# Market-Based Probability of Stock Returns

Victor Olkhov

Independent, Moscow, Russia

victor.olkhov@gmail.com

ORCID: 0000-0003-0944-5113

Jan. 30, 2026

**Abstract**

We derive the dependence of the market-based mean and variance of returns on the means, variances, and covariance of values and volumes of consecutive trades with shares of the security at the exchange. The conventional frequency-based definitions of means and variances describe only a limited case when all volumes of consecutive trades are assumed constant. To forecast market-based variance of returns at horizon *T*, one should predict the means, variances, and covariance of values and volumes of trades at the same horizon. We show that the correlation of returns and trade volumes is always zero. To estimate their statistical dependence, one should empirically calculate the correlation between returns and squares of trade volumes. We explain economic reasons that limit the number of predictable statistical moments of returns by the first two and the accuracy of the forecasts of the probability of returns by the accuracy of the Gaussian approximations. To adjust the reliability of large macroeconomic models like BlackRock's Aladdin, JP Morgan, and the U.S. Fed with economic reality that is determined by random trade volumes, the developers should use the market-based means and variances of returns.

This research received no support, specific grant or financial assistance from funding agencies in the public, commercial or nonprofit sectors. We welcome valuable offers of grants, support and positions.

## 1. Introduction

Studies of stock returns are endless (Ferreira and Santa-Clara, 2008; Diebold and Yilmaz, 2009; Kelly et al., 2022). The assessments of the factors that impact the expected return play a central role (Fisher and Lorie, 1964; Mandelbrot, Fisher, and Calvet, 1997; Campbell, 1985; Brown, 1989; Fama, 1990; Fama and French, 1992; Lettau and Ludvigson, 2003; Greenwood and Shleifer, 2013; Martin and Wagner, 2019). The irregular behavior of stock prices and returns makes probability theory a major tool for modeling returns. The probability distributions and correlation laws that may match the random return changes are studied by Kon (1984) and (Campbell, Grossman, and Wang, 1993; Llorente et al., 2001; Dorn, Huberman, and Sengmueller, 2008; Lochstoer and Muir, 2022). The description of the expected return is complemented by research on the realized return and volatility (Schlarbaum, Lewellen, and Lease, 1978; McAleer and Medeiros, 2008). The probability distributions of the realized and expected returns are studied by Amaral et al. (2000), Andersen et al., (2001), Knight and Satchell (2001), and Tsay (2005). That is only a tiny fraction of the returns' studies.

The market data are the basis for the empirical assessments of thee means and variances of returns. Fisher and Lorie (1964) considered "*Monthly closing prices of all common stocks on the NYSE from January 1926, through December 1960, have been placed on tape. .. Further, there are on tape - daily high, low, and closing prices of all common stocks on the NYSE from July 1, 1960, to the present.*" Lewellen, J. (1999) studied "*The empirical analysis focuses on industry portfolios formed monthly from May 1964 through December 1994, for a time series of 368 observations.*" Amaral et al. (2000) considered "*(i) the trades and quotes (TAQ) database, for which we analyze 40 million records for 1000 US companies for the 2-year period 1994-95, and (ii) the Center for Research and Security Prices (CRSP) database, for which we analyze 35 million daily records for approximately 16,000 companies in the 35-year period 1962-96. We study the probability distribution of returns over varying time scales from 5 min up to 4 years.*" Andersen et al. (2001) also stated that their "*empirical analysis is based on data from the Trade And Quotation (TAQ) database that contain continuously recorded information on the trades and quotations for the securities listed on the NYSE, AMEX, and the NASDAQ. Our sample extends from January 2, 1993, until May 29, 1998, for a total of 1,366 trading days.*" Nevertheless, "*the Trade And Quotation (TAQ) database*" provides time-stamped records of executed trades (prices and ***volumes***), Andersen et al. (2001) declared that their empirical studies of returns were formed on processing the time series of prices. For example, "..*The five-minute return series are constructed from the logarithmic difference between the prices recorded at or immediately before the corresponding five-minute marks.*"

All above studies use the time series of prices as the only source, which determines returns distributions. We show that the empirical studies of returns' means and variances, which processes only the time series of price records, have a limited use and may be incorrect and misleading while describing returns at real markets and exchanges. Such restricted empirical calculations of returns implicitly use a limited approximation of markets when all volumes of the consecutive trades with shares of all securities are covertly assumed constant. We present the market-based description of returns' variance that explicitly accounts for the impact of random volumes of consecutive trades and, as we derive, depends on the means, variances, and covariance of random values and volumes of consecutive trades.

In this paper we consider gross returns $R_g(t_i,\tau)$ (1.1) that are determined by the ratio of the payoff $x(t_i)$ (1.1) at time $t_i$ to price $p(t_i-\tau)$ in the past at time $t_i-\tau$:

$$R_g(t_i,\tau) = \frac{x(t_i)}{p(t_i-\tau)} = \frac{p(t_i)}{p(t_i-\tau)} + \frac{D(t_i)}{p(t_i-\tau)} \quad ; \quad x(t_i) = p(t_i) + D(t_i) \tag{1.1}$$

The payoffs $x(t_i)$ (1.1) are determined by two different random economic factors: by the trade prices $p(t_i)$ and dividends $D(t_i)$. We consider random trades with shares of the security as the first source that determines the market-based means and variances of returns $R(t_i,\tau)$ (1.2):

$$R(t_i,\tau) = \frac{p(t_i)}{p(t_i-\tau)} \quad ; \quad R_D(t_i,\tau) = \frac{D(t_i)}{p(t_i-\tau)} \quad ; \quad R_g(t_i,\tau) = R(t_i,\tau) + R_D(t_i,\tau) \tag{1.2}$$

The uncertain forecasts of the growth of the market value of the issuer that predict the payoffs of dividends $D(t_i)$ determine the second source of returns $R_D(t_i,\tau)$ (1.2) randomness. Actually, the highly frequent trades with shares of the security are executed at the exchange each working day. Contrary to that, the payoffs of the dividends as the result of the Board's decision are made usually once a year. We see no reasons for joint simultaneous consideration of these two completely different origins of the randomness of gross returns. The complexity of the relations that determine the means, variances, and covariance of both components of returns $R(t_i,\tau)$ and $R_D(t_i,\tau)$ (1.2) provide a basis for their separate studies. The obvious mutual dependence of both components of returns $R(t_i,\tau)$ and $R_D(t_i,\tau)$, especially close to time of the per-year dividend payoffs, should be studied at the second stage.

In this paper we study the market trade origin of random returns $R(t_i,\tau)$ (1.2) and present the market-based description of their means and variances. However, even the trade returns $R(t_i,\tau)$ (1.2) may be considered in two different cases.

As the first case of trade returns $R(t_i, \tau)$ (1.2), we regard the "portfolio returns" $R(t_i,t_0)$ like those described by Markowitz (1952). Let us view an investor who at time $t_0$ in the past purchased $U_0$ shares of a particular security. Since then, the investor hasn't traded his shares. At current time $t$, the investor observes the time series of trades executed at the exchange with shares of his security during a selected averaging interval $\Delta$ (1.3). The investor processes the time series of the observed trades at the exchange and assesses how the current trade statistics during $\Delta$ (1.3) model the current mean and variance of returns $R(t_i,t_0)$ of his shares with respect to their price $p(t_0)$ at time $t_0$ in the past. This case models the current market-based means and variances of share packages, which were purchased by the investors in the past (Sections 2 and 3).

The second case determines returns $R(t_i, \tau)$ (1.2) with fixed time shift $\tau$. This case was studied by Amaral et al. (2000) and Andersen et al. (2001). The averaging during the interval $\Delta$ (1.3) gives the assessments of means and variances of "imaginary" packages $U(t_i)$, which were purchased at time $t_i-\tau$ at price $p(t_i-\tau)$ and then sold at time $t_i$ at price $p(t_i)$. It is unlikely that this case describes the means and variances of any real share packages, but it may be useful for the investors and portfolio managers as an indicator of the dependence of the means and variances of actual trades with the time shift $\tau$. The description of the means and variance of returns $R(t_i, \tau)$ (1.2) differs from the means and variance of returns $R(t_i,t_0)$ that we indicate as the first case of "portfolio returns" and we describe returns $R(t_i, \tau)$ (1.2) in Sections 5.

We underline the different statistical dependence of the means and variances of returns $R(t_i,t_0)$ and $R(t_i, \tau)$. While presenting the results of empirical studies, one should always explicitly outline what kind of returns are under consideration.

Since Bachelier (1900), who highlighted the probabilistic nature of the price change, it has become routine to consider the probabilistic description of the fluctuations of prices and returns (Shiryaev, 1999; Shreve, 2004). We consider the random values and volumes of consecutive trades at the exchange as the origin of irregularity or stochasticity of returns during almost any period. To assess the mean and variance of returns, one should select the interval $\Delta$ (1.3) that contains $N$ terms of successive trades with shares and perform the averaging during $\Delta$.

$$\Delta = \left[t - \frac{\Delta}{2}; t + \frac{\Delta}{2}\right] \quad ; \quad t_i \in \Delta \quad ; \quad t_{i+1} = t_i + d \quad ; \quad i = 1,2,..N \tag{1.3}$$

The trades at the exchange are executed with high frequency, and the period between consecutive trades may be less than a second. The collecting and processing of time series of values and volumes of consecutive trades with such a high frequency is not too suitable for the description of the mean and variance of returns that are averaged during hours, days, or weeks. One may sum the values and volumes of all trades made at the exchange during a period $d$, which can be equal to minutes, hours, or days and obtain the time series of aggregated trades. Such a period $d$ (1.3) will determine the intervals between aggregated consecutive trades at the exchange, which are much more useful for their averaging during hours, days, or weeks, respectively. The number $N$ of consecutive trades during the averaging interval $\Delta$ (1.3), which may be equal to hours or days, should be sufficient for the reliable averaging of random returns determined by each trade at time $t_i$ during $\Delta$. For example, the number $N$ of trades during $\Delta$ may be $N \sim 50 - 100$. That determines the relations between the period $d$ (1.3) of successive trades and the averaging interval $\Delta$, and $d \ll \Delta$.

In Section 2, we introduce the return equation and market-based mean return. In Section 3, we derive market-based variance. In Section 4, we show that conventional frequency-based statistical moments of returns describe the case when all trade volumes are constant. In Section 5, we describe variance of return regarding the time shift $\tau$. Section 6 describes market-based return-volume correlations. In Section 7 we explain economic reasons that limit the number of predictable statistical moments of returns by the first two. Conclusion is in Section 8.

## 2. Market-based equation and mean return regarding the price $p(t_0)$ in the past

Let us consider the value $C(t_i)$, volume $U(t_i)$, and price $p(t_i)$ (2.1) of market trade executed at the exchange at time $t_i$ during the averaging interval $\Delta$ (1.3):

$$C(t_i) = p(t_i)U(t_i) \tag{2.1}$$

The randomness of portfolio returns $R(t_i,t_0)$ (2.2) of shares at time $t_i$ with respect to their price $p(t_0)$ in the past is completely determined by the random trade price $p(t_i)$ (2.1):

$$R(t_i, t_0) = \frac{p(t_i)}{p(t_0)} \tag{2.2}$$

The market-based mean, variance, and 3rd statistical moment of trade price $p(t_i)$ (2.1) were described in (Olkhov, 2022) and we refer there for details. Let us denote the *n-th* statistical moments of trade values $C(t;n)$ and volumes $U(t;n)$ averaged during $\Delta$ (1.3):

$$C(t;n) = E[C^n(t_i)] = \frac{1}{N}\sum_{i=1}^{N} C^n(t_i) \quad ; \quad C_\Sigma(t;n) = N \cdot C(t;n) \;\; ; \;\; n = 1,2,.. \tag{2.3}$$

$$U(t;n) = E[U^n(t_i)] = \frac{1}{N}\sum_{i=1}^{N} U^n(t_i) \quad ; \quad U_\Sigma(t;n) = N \cdot U(t;n) \tag{2.4}$$

We indicate that (2.3; 2.4) define the approximations of mathematical expectation $E[C^n(t_i)]$ and $E[U^n(t_i)]$ by a finite number $N$ of consecutive trades during $\Delta$ (1.3). Functions $C_\Sigma(t;n)$ and $U_\Sigma(t;n)$ denote the total sums of the n-th power of values $C^n(t_i)$ and volumes $U^n(t_i)$ of all $N$ trades during $\Delta$ (1.3). To highlight the difference of mathematical expectation $E[..]$ that defines the *n-th* statistical moments of trade values $C(t;n)$ (2.3) and volumes $U(t;n)$ (2.4) and market-based mathematical expectation of prices and returns we denote it $E_m[..]$. We propose that the past price $p(t_0)$ of shares is known, and hence, the mean return $R(t,t_0;1)$ (2.5) and variance $\Theta(t)$ (2.6) correspond to the mean $p(t;1)$ and variance $\Phi(t)$ of random prices $p(t_i)$ (2.1):

$$R(t,t_0;1) = E_m[R(t_i,t_0)] = \frac{E_m[p(t_i)]}{p(t_0)} = \frac{p(t;1)}{p(t_0)} \quad ; \quad p(t;1) = E_m[p(t_i)] \qquad (2.5)$$

$$\Theta(t) = E_m[(R(t_i,t_0) - R(t;1))^2] = \frac{E_m[(p(t_i)-p(t;1))^2]}{p^2(t_0)} = \frac{\Phi(t)}{p^2(t_0)} \qquad (2.6)$$

$$\Phi(t) = E_m[(p(t_i) - p(t;1))^2] \qquad (2.7)$$

We take the market-based mean price $p(t;1)$ (2.1; 2.5) as volume weighted average price (VWAP) (Berkowitz et al., 1988; Duffie and Dworczak, 2018) that takes the form:

$$p(t;1) = E_m[p(t_i)] = \frac{1}{U_\Sigma(t;1)}\sum_{i=1}^{N} p(t_i)\cdot U(t_i) = \frac{C(t;1)}{U(t;1)} = \frac{C_\Sigma(t;1)}{U_\Sigma(t;1)} \qquad (2.8)$$

The relations (2.5) and (2.8) define market-based mean return $R(t,t_0;1)$ (2.9):

$$R(t,t_0;1) = E_m[R(t_i,t_0)] = \frac{E_m[p(t_i)]}{p(t_0)} = \frac{1}{p(t_0)U_\Sigma(t;1)}\sum_{i=1}^{N} \frac{p(t_i)}{p(t_0)}\, p(t_0)\cdot U(t_i) \qquad (2.9)$$

We introduce the functions $C_\tau(t_i,t_0)$ (2.10) that have the meaning of the past value of the current trade volumes $U(t_i)$ at price $p(t_0)$ in the past at time $t_0$.

$$C_\tau(t_i,t_0) = p(t_0)\cdot U(t_i) \quad ; \quad C_\tau(t,t_0;n) = E[C_\tau^{\,n}(t_i;t_0)] = \frac{1}{N}\sum_{i=1}^{N} C_\tau^{\,n}(t_i;t_0) \qquad (2.10)$$

$$C_{\tau\Sigma}(t,t_0;n) = \sum_{i=1}^{N} C_\tau^{\,n}(t_i;t_0) \qquad (2.11)$$

To underline, that $C_\tau(t_i,t_0)$ (2.10) denotes the value with a time shift, we use index $\tau$. We define the n-th statistical moments $C_\tau(t,t_0;n)$ of the past random values $C_\tau(t_i,t_0)$ (2.10) like (2.3), and $C_{\tau\Sigma}(t,t_0;n)$ (2.11) determine the total sum of the n-th power of the past random values $C_\tau^{\,n}(t_i,t_0)$ during $\Delta$ (1.3). The use of (2.10; 2.11) presents the mean return $R(t,t_0;1)$ (2.9) as (2.12):

$$R(t,t_0;1) = \frac{1}{C_{\tau\Sigma}(t,t_0;1)}\sum_{i=1}^{N} R(t_i,t_0)\cdot C_\tau(t_i,t_0) = \frac{C(t;1)}{C_\tau(t,t_0;1)} = \frac{C_\Sigma(t;1)}{C_{\tau\Sigma}(t,t_0;1)} \qquad (2.12)$$

The expression of market-based mean return $R(t,t_0;1)$ as (2.9) as the consequence of VWAP $p(t;1)$ (2.8) almost completely reproduces the definition of mean return (2.13) of the portfolio that was given by Markowitz (1952):

$$R(t,t_0;1) = \sum_{i=1}^{N} R(t_i,t_0)\cdot X_i(t) \quad ; \quad X_i(t) = \frac{C_\tau(t_i,t_0)}{C_{\tau\Sigma}(t,t_0;1)} = \frac{p(t_0)\cdot U(t_i)}{p(t_0)\cdot U_\Sigma(t;1)} = \frac{U(t_i)}{U_\Sigma(t;1)} \qquad (2.13)$$

Markowitz (1952) considered a portfolio composed by $j=1,2,..J$ securities in the past at time $t_0$ and denoted $X_j(t_0)$ as the relative amounts invested into security $j$. The market-based mean return $R(t,t_0;1)$ as (2.9; 2.12) is like Markowitz' mean return (2.13), but the functions $X_i(t)$ have the meaning of past values $C_\tau(t_i,t_0)$ (2.10) of current trade volumes $U(t_i)$ at price $p(t_0)$ in the past at time $t_0$. One may consider the past values $C_\tau(t_i,t_0)$ (2.10) determined by the trade volumes $U(t_i)$ at time $t_i$, $i=1,..N$, during $\Delta$ (1.3) as the values of "imaginary security $i$" of the portfolio in the past at time $t_0$. These use of "imaginary securities $i$" establishes almost total similarity with Markowitz's expression of the portfolio mean return. We underline the similarities between market-based mean return $R(t,t_0;1)$ (2.12) and Markowitz's mean return (2.13) to outline that Markowitz actually described value weighted average returns, and due to parallels (2.8; 2.9; 2.12) and (2.13), one may consider it as the expression of volume weighted average price (2.8), presented by Markowitz 35 years ahead of Berkowitz et al. (1988).

The equations (2.2; 2.10) present the random price $p(t_i)$ equation (2.1) as the equation on random returns $R(t_i,t_0)$ (2.14):

$$C(t_i) = p(t_i)U(t_i) = \frac{p(t_i)}{p(t_0)} \cdot p(t_0)U(t_i) = R(t_i,t_0) \cdot C_\tau(t_i,t_0)$$

$$C(t_i) = R(t_i,t_0) \cdot C_\tau(t_i,t_0) \tag{2.14}$$

The equation (2.14) ties up the current trade value $C(t_i)$ (2.1) of the trade volume $U(t_i)$ (2.1) at current price $p(t_i)$ (2.1) and the past value $C_\tau(t_i,t_0)$ (2.10) of the same trade volume $U(t_i)$ at price $p(t_0)$ in the past at time $t_0$. The current trade values $C(t_i)$ (2.1) and the past values $C_\tau(t_i,t_0)$ (2.10) determine the returns $R(t_i,t_0)$ (2.2; 2.14) with respect to the price $p(t_0)$ in the past at time $t_0$. It is obvious that the equation (2.14) has the same form as (2.1). Hence, statistical moments of returns $R(t_i,t_0)$ (2.14) depend on statistical moments of the current values $C(t_i)$ (2.1) and the past values $C_\tau(t_i,t_0)$ (2.10) in the same form as statistical moments of prices $p(t_i)$ (2.1) depend on statistical moments of the current values $C(t_i)$ and volumes $U(t_i)$ (2.1).

## 3. Market-based variance regarding the price $p(t_0)$ in the past

The iterative relations (3.1) determine the market-based *n-th* statistical moment of price $p(t;n)=E_m[p^n(t_i)]$ through the previous statistical moments $p(t;m)$, $m=1,..n-1$ (Olkhov, 2022), and we refer there for further details.

$$E_m[p^n(t_i)] = E_m[p(t_i)p^{n-1}(t_i)] = E_m[p(t_i)]E_m[p^{n-1}(t_i)] + cov[p(t_i),p^{n-1}(t_i)] \tag{3.1}$$

The *n-th* power of equation (2.1), relations (2.4), and VWAP $p(t;1)$ (2.9) define the equations:

$$C^n(t_i) = p^n(t_i) \cdot U^n(t_i) \quad ; \quad w(t_i;n) = \frac{U^n(t_i)}{U_\Sigma(t;n)} \quad ; \quad \sum_{i=1}^{N} w(t_i;n) = 1 \tag{3.2}$$

The averaging by the weight function $w(t_i;n)$ determines the covariance $cov[p(t_i),p^{n-1}(t_i)]$ (3.3):

$$cov[p(t_i),p^{n-1}(t_i)] = E_m[(p(t_i) - p(t;1)(p^{n-1}(t_i) - p(t;n-1)]$$

$$cov[p(t_i),p^{n-1}(t_i)] = \frac{1}{N}\sum_{i=1}^{N}\big(p(t_i) - p(t;1)\big)\big(p^{n-1}(t_i) - p(t;n-1)\big)\, w(t_i;n) \tag{3.3}$$

For $n=2$, the relations (3.1-3.3) define market-based variance $\Phi(t)$ (3.4) of prices and the 2nd statistical moment $p(t;2)$ (3.4) (Olkhov, 2022):

$$\Phi(t) = E_m[\big(p(t_i) - p(t;1)\big)^2]$$

$$\Phi(t) = \frac{\psi^2(C) - 2\,\varphi(C,U) + \chi^2(U)}{1+\chi^2(U)} \cdot p^2(t;1) \quad ; \quad p(t;2) = [1 + \frac{\psi^2(C) - 2\,\varphi(C,U) + \chi^2(U)}{1+\chi^2(U)}] \cdot p^2(t;1) \tag{3.4}$$

The market-based variance $\Phi(t)$ (3.4) of price depends on the coefficient of variation $\psi(C)$ of the trade values $C(t_i)$, on the coefficient of variation $\chi(U)$ of the trade volumes $U(t_i)$, and on their covariance $\varphi(C,U)$ (3.5):

$$\psi^2(C) = \frac{cov[C(t_i),C(t_i)]}{C^2(t;1)} \quad ; \quad \chi^2(U) = \frac{cov[U(t_i),U(t_i)]}{U^2(t;1)} \quad ; \quad \varphi(C,U) = \frac{cov[C(t_i),U(t_i)]}{C(t;1)U(t;1)} \tag{3.5}$$

We use (3.5) to transform $C(t;2)$ (2.3), $U(t;2)$ (2.4), and the covariance $cov[C(t_i),U(t_i)]$ as:

$$C(t;2) = \left(1+\psi^2(C)\right)C^2(t;1) \quad ; \quad U(t;2) = (1+\chi^2(U))U^2(t;1) \tag{3.6}$$

$$cov[C(t_i),U(t_i)] = \frac{1}{N}\sum_{i=1}^{N}[C(t_i)-C(t;1)][U(t_i)-U(t;1)] = \varphi(C,U)C(t;1)U(t;1) \tag{3.7}$$

The market-based variance $\Phi(t)$ (3.4) determines the variance $\Theta(t,t_0)$ (3.8; 3.9) of returns:

$$\Theta(t,t_0) = E_m\left[\left(R(t_i,t_0)-R(t,t_0;1)\right)^2\right] \tag{3.8}$$

$$\Theta(t,t_0) = \frac{\Phi(t)}{p^2(t_0)} = \frac{\psi^2(C)-2\,\varphi(C,U)+\chi^2(U)}{1+\chi^2(U)}\cdot R^2(t,t_0;1) \tag{3.9}$$

From (3.8; 3.9), obtain market-based 2nd statistical moment of return $R(t,t_0;2)$ (3.10):

$$R(t,t_0;2) = E_m[R^2(t_i,t_0)] = \frac{E_m[p^2(t_i)]}{p^2(t_0)} = \left[1+\frac{\psi^2(C)-2\,\varphi(C,U)+\chi^2(U)}{1+\chi^2(U)}\right]\cdot R^2(t,t_0;1) \tag{3.10}$$

One may obtain the same market-based variance $\Theta(t,t_0)$ (3.8) and the 2nd statistical moment of return $R(t,t_0;2)$ (3.10) using the relations like (3.2). Indeed, the n-th power of the equation (2.14) and market-based mean return (2.12) cause equations (3.11) like the equations (3.2):

$$C^n(t_i) = R^n(t_i,t_0)\cdot {C_\tau}^n(t_i,t_0) \quad ; \quad z(t_i,t_0;n) = \frac{{C_\tau}^n(t_i,t_0)}{C_{\tau\Sigma}(t,t_0;n)} \quad ; \quad \sum_{i=1}^{N} z(t_i,t_0;n) = 1 \tag{3.11}$$

From (3.11) for $n=1$ obtain the equation (2.14) and mean $R(t,t_0;1)$ (2.12). For $n=2$, obtain:

$$C^2(t_i) = R^2(t_i,t_0)\cdot {C_\tau}^2(t_i,t_0) \tag{3.12}$$

Similar to (3.1; 3.3; 3.4), obtain:

$$R(t,t_0;2) = E_m[R^2(t_i,t_0)] = {E_m}^2[R(t_i,t_0)] + E_m\left[\left(R(t_i,t_0)-R(t,t_0;1)\right)^2\right] \tag{3.13}$$

$$\Theta(t,t_0) = E_m\left[\left(R(t_i,t_0)-R(t,t_0;1)\right)^2\right] = \frac{1}{N}\sum_{i=1}^{N}\left(R(t_i,t_0)-R(t,t_0;1)\right)^2 z(t_i,t_0;2) \tag{3.14}$$

The equation (3.14) gives (3.8; 3.9), and (3.13) gives (3.10).

One may use the relations (3.1-3.3) for $n=3$ and derive the market-based Skewness $Sk_m(R)$ and 3rd statistical moment of returns or use the expressions of price market-based Skewness $Sk_m(p)$ (Olkhov, 2022) and the equality of price and return Skewness: $Sk_m(R) = Sk_m(p)$.

**4. Statistical moments of returns when all trade volumes are constant**

We show that for a limited approximation, which assumes that all volumes of consecutive trades are constant, the statistical moments of prices and returns take the usual frequency-based expressions that are like (2.3; 2.4). Let us consider the *n-th* statistical moments of values $C(t;n)$ (2.3), the equations (3.2), and put all trade volumes $U(t_i)=U$ constant during $\Delta$ (1.3):

$$C(t;n)=E[C^n(t_i)]=\frac{1}{N}\sum_{i=1}^{N}C^n(t_i)=\frac{1}{N}\sum_{i=1}^{N}p^n(t_i)\cdot U^n(t_i)=U^n\frac{1}{N}\sum_{i=1}^{N}p^n(t_i) \quad (4.1)$$

We denote $\pi(t;n)$ frequency-based statistical moments of prices:

$$\pi(t;n)=E[p^n(t_i)]=\frac{1}{N}\sum_{i=1}^{N}p^n(t_i)=\frac{E[C^n(t_i)]}{U^n}=\frac{C(t;n)}{U^n} \quad (4.2)$$

The relations (4.2) define conventional expressions of frequency-based statistical moments of prices $\pi(t;n)$. One should keep in mind that these assessments describe statistical moments $\pi(t;n)$ (4.2) in the approximation when all volumes of consecutive trades are constant. Otherwise, one should consider the market-based mean price $p(t;1)$ (2.8), variance of price $\Phi(t)$ (3.4), and higher market-based statistical moments.

The same applies to statistical moments of returns $R(t,t_0;n)$. From (2.3), (3.11), and the definition of $C_\tau(t_i,t_0)$ (2.10), for the case of constant trade volumes $U(t_i)=U$ during $\Delta$ (1.3):

$$C(t;n)=E[C^n(t_i)]=\frac{1}{N}\sum_{i=1}^{N}C^n(t_i)=\frac{1}{N}\sum_{i=1}^{N}R^n(t_i,t_0)C_\tau{}^n(t_i,t_0)=p^n(t_0)U^n\frac{1}{N}\sum_{i=1}^{N}R^n(t_i,t_0)$$

From above, obtain:

$$\rho(t,t_0;n)=E[R^n(t_i,t_0)]=\frac{1}{N}\sum_{i=1}^{N}R^n(t_i,t_0)=\frac{C(t;n)}{p^n(t_0)U^n} \quad (4.3)$$

In (4.3), at current time $t$, we denote $\rho(t,t_0;n)$ the assessment of the *n-th* statistical moment of returns of shares that were purchased at time $t_0$ in the past. The statistical moments $\rho(t,t_0;n)$ (4.3) or returns with respect price $p(t_0)$ in the past describe the case when all volumes of consecutive trades are assumed constant during $\Delta$ (1.3). Otherwise, one should calculate the market-based mean return $R(t,t_0;1)$ (2.12), the variance $\Theta(t,t_0)$ (3.8; 3.9) of returns, etc.

The use of frequency-based statistical moments of prices $\pi(t;n)$ (4.2) or returns $\rho(t,t_0;n)$ (4.3) implicitly assumes the approximation of constant trade volumes. Meanwhile, the time series of consecutive trade volumes at real financial markets and exchanges are highly irregular or random. The implicit use of a limited approximation of constant trade volumes and relations $\pi(t;n)$ (4.2) or $\rho(t,t_0;n)$ (4.3) may result in a low accuracy and wrong forecasts of the corresponding economic and financial models.

## 5. The mean return and variance regarding the time shift $\tau$

In this Section we consider the description of the mean and variance for the second case of returns $R(t_i,\tau)$ (1.2), which accounts for the dependence on the time shift $\tau$. Such a case was studied, in particular, by Amaral et al. (2000), who described returns with respect to "*a wide range of time scales $\tau$, varying over 3 orders of magnitude, 5 min≤$\tau$≤6240 min= 16 days*."

However, as others, Amaral et al. (2000) used only historical time series of market prices and ignored the impact of the corresponding trade volumes. Their derivations of histograms and probability distributions of returns (Amaral et al., 2000) are implicitly based on the assumption constant trade volumes. Moreover, neglecting random volumes of consecutive trades while describing returns that actually depend on random trade volumes raises great doubts in the economic sense and definitely in the accuracy of the results. We propose that this and similar studies should be reperformed to clarify the impact of random trade volumes.

To describe market-based mean, variance and higher statistical moments of returns $R(t_i, \tau)$ (1.2), one should follow equations (2.10; 3.11) but account for the dependence on the time shift $\tau$ instead of the dependence on the price $p(t_0)$ in the past at time $t_0$. We highlight that we consider market trade returns $R(t_i, \tau)$ (1.2) that are determined by the values and volumes of market trades at the exchange and don't account for the payoffs of dividends by the issuers. We introduce the equations that account for the time shift $\tau$.

$$C_\tau(t_i,\tau) = p(t_i - \tau)\cdot U(t_i) \quad ; \quad C_\tau(t,\tau;n) = E[{C_\tau}^n(t_i;\tau)] = \frac{1}{N}\sum_{i=1}^{N} {C_\tau}^n(t_i;\tau) \tag{5.1}$$

We call functions $C_\tau(t_i, \tau)$ (5.1) past values of current trade volume $U(t_i)$ at price $p(t_i-\tau)$.

$$R(t_i,\tau) = \frac{p(t_i)}{p(t_i-\tau)} \quad ; \quad C_{\tau\Sigma}(t,\tau;n) = \sum_{i=1}^{N} {C_\tau}^n(t_i;\tau) \tag{5.2}$$

The relations (5.2) define return $R(t_i, \tau)$ of trade at time $t_i$ at price $p(t_i)$ with respect to the price $p(t_i-\tau)$ at time $t_i-\tau$. The functions $C_{\tau\Sigma}(t,\tau;n)$ denote the total sum of the n-th power $C_\tau^n(t_i, \tau)$ of past values during the averaging interval $\Delta$ (1.3) and they are similar to $C_\Sigma(t;n)$ (2.11).

$$C^n(t_i) = R^n(t_i,\tau)\cdot {C_\tau}^n(t_i,\tau) \quad ; \quad z(t_i,\tau;n) = \frac{{C_\tau}^n(t_i,\tau)}{C_{\tau\Sigma}(t,\tau;n)} \quad ; \quad \sum_{i=1}^{N} z(t_i,\tau;n) = 1 \tag{5.3}$$

The equations (5.3) have the form like the equations (3.2) and (3.11). The market-based mean return $R(t,\tau;1)$ (5.4) takes the form like (2.8; 2.9):

$$R(t,\tau;1) = \frac{1}{C_{\tau\Sigma}(t,\tau;1)}\sum_{i=1}^{N} R(t_i,\tau)\cdot C_\tau(t_i,\tau) = \frac{C(t;1)}{C_\tau(t,\tau;1)} = \frac{C_\Sigma(t;1)}{C_{\tau\Sigma}(t,\tau;1)} \tag{5.4}$$

The important comment: the market-based mean return $R(t,\tau;1)$ (5.4) reveals the dependence on the random volumes of consecutive trades $U(t_i)$ during the averaging interval $\Delta$ (1.3) and on random prices $p(t_i-\tau)$ during the averaging interval $\Delta_\tau$ (5.5):

$$\Delta_\tau = \left[(t-\tau)-\frac{\Delta}{2};(t-\tau)+\frac{\Delta}{2}\right] \quad ; \quad t_i - \tau \in \Delta_\tau \quad ; \quad t_{i+1} = t_i + d \quad ; \quad i = 1,2,..N \tag{5.5}$$

The market-based variance $\Theta(t,\tau)$ (5.6) of returns takes the form similar to (3.9):

$$\Theta(t,\tau) = \frac{\psi^2(C) - 2\,\varphi(C,C_\tau) + \chi^2(C_\tau)}{1+\chi^2(C_\tau)}\cdot R^2(t,\tau;1) \tag{5.6}$$

The functions that define market-based variance $\Theta(t,\tau)$ (5.6) have the meaning of the coefficient of variation $\chi(C_\tau)$ of the past values of the current trade volumes $U(t_i)$ at price $p(t_i-$ $\tau)$ with time shift $\tau$, and of the covariance $\varphi(C, C_\tau)$ of current trade values $C(t_i)$ and past values $C_\tau(t_i, \tau)$ (5.1) with time shift $\tau$:

$$\chi^2(C_\tau) = \frac{cov[C_\tau(t_i,\tau), C_\tau(t_i,\tau)]}{{C_\tau}^2(t,\tau;1)} \quad ; \quad \varphi(C, C_\tau) = \frac{cov[C(t_i), C_\tau(t_i,\tau)]}{C(t;1)C_\tau(t,\tau;1)} \tag{5.7}$$

The higher market-based *n-th* statistical moments of returns $R(t, \tau;n)$ follow from the equations similar to (3.1-3.3) with the averaging over the weight functions $z(t_i, \tau;n)$ (5.3).

The conventional frequency-based statistical moments of return $\rho(t, \tau;n)$ result from (2.3; 5.3) the limited case when all volumes of consecutive trades $U(t_i)$ and the prices $p(t_i-\tau)$ at time $t_i-\tau$ are assumed constant. That causes the values $C_\tau(t_i, \tau)$ (5.1) of trade volumes $U(t_i)$ at price $p(t_i-\tau)$ with time shift $\tau$ are constant for all $t_i$ during $\Delta$ (1.3). This approximation coincides with the case that describes frequency-based statistical moments of returns $\rho(t,t_0;n)$ (4.3) regarding the price $p(t_0)$ when all trade volumes are constant.

## 6. Market-Based Return-Volume Correlations

The investors and portfolio managers observe statistical dependence between volumes of trades and returns. However, the empirical calculations of statistical dependence (Campbell et al., 1993; Llorente et al., 2001) between time series of trade volumes and returns, at best, use the approximation of constant trade volumes, which is rather inconsistent with the goals to empirically estimate the correlations with variable trade volumes. Moreover, one should keep in mind that returns are not directly observed. Researchers derive historical time series of returns by processing the historical time series of prices. The different manipulations with the time series of prices of the executed trades result in time series of returns that may have different statistical dependence on trade volumes. We highlight that we consider returns that are generated by market trades and don't account for the impact of the dividends that may occur once a year. However, even in this case, the processing of the time series of prices may determine the time series of returns that show at least two types of statistical dependence on trade volumes. We have in mind the time series of returns $R(t_i,t_0)$ (2.2) regarding the fixed price $p(t_0)$ at time $t_0$ in the past and returns $R(t_i, \tau)$ (1.2) with respect to the price $p(t_i-\tau)$ with the time shift $\tau$. The first case initially was described by Markowitz (1952), and we may refer to it as "portfolio returns." The second one was empirically studied, for example, by Amaral et al. (2000). We recall that the different definitions of returns result in the different means, variances, and different statistical dependence on trade volumes.

We highlight that two time series of economic variables don't determine themselves the method for the assessments of their correlation. Any empirical correlation between two economic time series should always follow a particular economic approximation. Even the simplest conventional frequency-based correlation of two economic variables implicitly uses a particular economic approximation. Below we consider two different correlations. The first one determines the correlation between "portfolio returns" $R(t_i,t_0)$ (2.2) with respect to the price $p(t_0)$ at time $t_0$ in the past and trade volumes. The second one describes the correlation between returns $R(t_i,\tau)$ (1.2) with respect to the price $p(t_i-\tau)$ with the time shift $\tau$ and the values $C_\tau(t_i,t_0)$ (2.10) with a time shift $\tau$.

Both cases explicitly account for the impact of random trade volumes, and we argue the limitations of the constant trade volumes approximation.

***6.1 Market-based correlation between "portfolio returns" $R(t_i,t_0)$ and trade volumes $U(t_i)$***

The market-based correlation during the averaging interval $\Delta$ (1.3) between the time series of volumes $U(t_i)$ of consecutive trades and returns $R(t_i,t_0)$ (2.2) with respect to price $p(t_0)$ in the past is determined by the correlation of trade volumes $U(t_i)$ and prices $p(t_i)$ and should obey:

$$cov[R(t_i,t_0),U(t_i)] = \frac{E_m[(p(t_i)-E_m[p(t_i)])\cdot(U(t_i)-E_m[U(t_i)])]}{p(t_0)} = \frac{cov[p(t_i),U(t_i)]}{p(t_0)}$$

$$cov[p(t_i),U(t_i)] = E_m[(p(t_i)-E_m[p(t_i)])\cdot(U(t_i)-E_m[U(t_i)])] \qquad (6.1)$$

The definition (6.1), (2.1-2.4), and (2.8) give:

$$cov[p(t_i),U(t_i)] = E_m[p(t_i)U(t_i)] - E_m[p(t_i)]E_m[U(t_i)] = E_m[C(t_i)] - p(t;1)U(t;1) \qquad (6.2)$$

The use of (2.1; 2.3; 2.4; 2.8), give:

$$C(t;1) = \frac{1}{N}\sum_{i=1}^{N} p(t_i)U(t_i) = \frac{1}{\sum_{i=1}^{N} U(t_i)}\sum_{i=1}^{N} p(t_i)U(t_i)\cdot\frac{1}{N}\sum_{i=1}^{N} U(t_i) = p(t;1)U(t;1)$$

Hence, from (6.1; 6.2), obtain that correlation $cov[R(t_i,t_o),U(t_i)]$ (6.1) always equals zero:

$$cov[R(t_i,t_0),U(t_i)] = \frac{cov[p(t_i),U(t_i)]}{p(t_0)} = \frac{C(t;1)-p(t;1)U(t;1)}{p(t_0)} = 0 \qquad (6.3)$$

Meanwhile, zero correlation $cov[R(t_i,t_o),U(t_i)]$ between returns $R(t_i,t_0)$ (2.2) and trade volumes $U(t_i)$ doesn't results in the absence of their statistical dependence. Those, who empirically study statistical dependence should calculate the correlation $cov[R(t_i,t_o),U^2(t_i)]$ between returns $R(t_i,t_0)$ (2.2) and squares of trade volumes $U^2(t_i)$:

$$cov[R(t_i,t_0),U^2(t_i)] = E_m[R(t_i,t_0)U^2(t_i)] - E_m[R(t_i,t_0)]E_m[U^2(t_i)] \qquad (6.4)$$

From (2.1; 2.2), obtain:

$$E_m[R(t_i,t_0)U^2(t_i)] = \frac{E_m[p(t_i)U^2(t_i)]}{p(t_0)} = \frac{E_m[C(t_i)U(t_i)]}{p(t_0)}$$

The market-based means of returns (2.8) and means of squares of volumes (2.4):

$$E_m[R(t_i,t_0)] = R(t,t_0;1) = \frac{p(t;1)}{p(t_0)} \quad ; \quad E_m[U^2(t_i)] = U(t;2)$$

The joint mean $E_m[C(t_i),U(t_i)]$ of trade values $C(t_i)$ and volumes $U(t_i)$ has the usual form:

$$E_m[C(t_i)U(t_i)] = \frac{1}{N}\sum_{i=1}^{N} C(t_i)U(t_i) \quad (6.5)$$

Finally, obtain the expression of $cov[R(t_i,t_o),U^2(t_i)]$:

$$cov[R(t_i,t_0),U^2(t_i)] = \frac{1}{p(t_0)\cdot N}\sum_{i=1}^{N} C(t_i)U(t_i) - R(t,t_0;1)\cdot U(t;2) \quad (6.6)$$

We underline the "formal" reason that the empirical calculations of a "conventional" definition of the correlation between returns and trade volumes are completely improper. Indeed, "conventional" definition of $cov_{emp}[R(t_i,t_o),U(t_i)]$ is follows:

$$cov_{emp}[R(t_i,t_0),U(t_i)] = \frac{1}{N}\sum_{i=1}^{N}(R(t_i,t_0) - E_m[R(t_i,t_0)])\cdot(U(t_i) - E_m[U(t_i)]) \quad (6.7)$$

However, this equation gives:

$$cov_{emp}[R(t_i,t_0),U(t_i)] = \frac{1}{N}\sum_{i=1}^{N} R(t_i,t_0)\,U(t_i) - E_m[U(t_i)]\cdot\frac{1}{N}\sum_{i=1}^{N} R(t_i,t_0) \quad (6.8)$$

From (4.3) and (6.8), obtain, that the use of "conventional" expression of the correlation $cov_{emp}[R(t_i,t_o),U(t_i)]$ (6.7) between returns $R(t_i,t_o)$ (2.2) and trade volumes $U(t_i)$ is based implicitly on the use of the mean return $\rho(t,t_0;1)$ (4.3; 6.9), and hence, describes only a limited cased when volumes of consecutive trades are constant $U(t_i)=U$ during the interval $\Delta$ (1.3).

$$\rho(t,t_0;1) = \frac{1}{N}\sum_{i=1}^{N} R(t_i,t_0) \quad (6.9)$$

Meanwhile, if all trade volumes are constant $U(t_i)=U$, then any correlations with time series of constant trade volumes are zero. One may use (2.1-2.12) and (4.2;4.2) to show that the "empirical calculations" of correlation $cov_{emp}[R(t_i,t_o),U(t_i)]$ (6.7) actually estimate the difference between market-based mean return $R(t,t_0;1)$ (2.8; 2.9) and the mean return $\rho(t,t_0;1)$ (4.3; 6.9) when all trade volumes are constant during $\Delta$ (1.3):

$$cov_{emp}[R(t_i,t_0),U(t_i)] = [R(t,t_0;1) - \rho(t,t_0;1)]U(t;1) = \frac{p(t;1) - \pi(t;1)}{p(t_0)}U(t;1) \quad (6.10)$$

### ***6.2 Market-based correlation between returns $R(t_i,\tau)$ and past values $C_\tau(t_i,\tau)$***

The assessment of market-based correlation of returns $R(t_i,\tau)$ (5.2) with respect to the price $p(t_i-\tau)$ with the time shift $\tau$ and past trade values $C_\tau(t_i,\tau)$ (5.1) completely similar to market-based description of correlation between returns $R(t_i,t_0)$ (2.2) with respect to price $p(t_0)$ in the past and the trade volumes $U(t_i)$, which we described in the subsection 6.1. One may simply substitute the notations $R(t_i,t_0)$ by $R(t_i,\tau)$, and notations $U(t_i)$ by $C_\tau(t_i,\tau)$ to obtain the similar relations.

However, the market-based assessment of correlation between returns $R(t_i,\tau)$ (5.2) and trade volumes $U(t_i)$ is a rather complex problem, and we don't consider it in this paper.

## 7. Economic complexity limits the number of predictable statistical moments

In this section we briefly explain the economic reasons that limit the number of predictable market-based statistical moments of returns, at best, by the first two.

We derived that the market-based means and variances of prices and returns of a particular security *A* depend on the means, variances, and covariances of values and volumes of consecutive trades at the exchange with shares of this security. In turn, the trades with security *A* considerably depend, on the one hand, on the market trades with securities of the same industry and, on the other hand, on trades of the issuer *A* in the relevant markets, which determine the economic performance and growth of the issuer *A*. The trades at the exchange with securities of a particular industry sector, in its turn, depend on the performance of the market portfolio, on the trades with all securities at the exchange. Ultimately, the means and variances of prices and returns of the market portfolio are determined by the means, variances and covariance of values and volumes of trades made with shares of all traded securities. Meanwhile, the trade with shares at the exchange is only one of many sides of the economic trade as a whole, which should account for the values and volumes of OTC, trades with commodities, energy, bank credit and transfers, retail, etc., etc. Economic growth and development during a particular time interval *D* is determined by the set of values and volumes of economic and financial trades of a great diversity. It is evident that trades with a particular security, or securities of a particular industry sector, or trades with all securities of the market portfolio, are only pieces of the general trades executed in the economy during some time period *D*. The changes of macroeconomic variables such as investment and consumption, credit and sales, etc., during the period *D* are completely determined by the values and volumes of corresponding transaction made in numerous markets of the economy.

This permits to conclude that means, variances, and covariances of values and volumes of financial and economic trades in the economy as a whole completely determine means and variances of prices and returns, and moreover, determine means and variances of macroeconomic variables (Olkhov, 2021; 2023a; 2023b; 2024). The accuracy of observations of most macroeconomic variables, which have meaning of linear forms of means of values or volumes of economic transactions, is limited by the accuracy of observations of trades, variances and covariances of their trade values and volumes. However, the total records of time series of values and volumes of all economic and financial transactions during the period *D*, are absent. Current econometric methodologies (Fox et al., 2019) use available data of economic transactions or substitute them by other economic measurements to estimate macroeconomic variables. These challenges additionally increase the uncertainty and reduce the accuracy of econometric assessments.

We denote macroeconomic variables that are determined as linear forms of means of values or volumes of economic transactions, as the 1st order variables. We call macroeconomic variables determined by quadratic forms of variances and covariances of trade values or volumes as the 2nd order variables, and etc. Actually, current econometrics (Fox et al., 2019) assesses mostly macroeconomic variables of the 1st order. One may mention the covariances of prices and returns of valuable securities, or industries as almost only 2nd order variables that are estimated by econometrics. That reveals a tough economic problem, which reduces the accuracy of the predictions of prices, returns, and macroeconomic variables.

On the one hand, economic theories use econometric assessments to forecast the macroeconomic variables, which mostly have meaning of the 1st order. Only few 2nd order variables, like covariances of most valuable prices and returns, are taken into account in macroeconomic models and forecasts. That is completely insufficient for economic-based predictions of covariances of values and volumes of the most part of economic and financial transactions executed in the economy during the period *D*. Modern econometrics (Fox et al., 2019) has no observations of values and volumes of most consecutive transactions in the economy and can't provide the assessments of their covariances.

Finally, the lack of econometric observations and assessments of covariances of values and volumes of most economic and financial transactions in the economy makes impossible any economic-based predictions of variances and covariances of trades at any reasonable horizon *T*. The lack of econometric assessments of variances of economic transactions limits the number of predictable market-based statistical moments by the first two. That limits the accuracy of any forecasts of macroeconomic variables, prices, returns, volumes and values of economic and financial transactions by the accuracy of Gaussian approximations.

Modern economic theories can't describe mutual dependence of the 1st and 2nd order variables. That makes it impossible the economic-based forecasts of 2nd statistical moments of trade values and volumes, prices, and returns. One may use historical trade records to assess the current 3rd and 4th statistical moments of prices and returns like Skewness or Kurtosis. However, there is no economic ground for their predictions at horizon *T*. The nearest, but still far goal is the economic-based description of the 1st and 2nd order variables. Till then, the economic-founded predictions of the market-based statistical moments of trades, prices, returns, and macroeconomic variables are limited by the first two.

## 8. Conclusion

We present the market-based dependence of means and variances of returns on the means, variances, and covariances of values and volumes of consecutive trades at exchange. We propose the rules for the derivation of higher market-based statistical moments of returns, but don't provide the pure prove. At modern exchanges the consecutive trades are executed with period less than a second and payoffs of dividends occur once-a-year. The big difference in the frequences of these events allows describe the trade origin of return's variance without accounting for impact of dividend payoffs.

We show that market-based mean and variance of returns take their usual frequency-based forms if all volumes of the consecutive trades with the security at exchange are assumed constant during the averaging interval. If the coefficient of variation of trade volumes is sufficiently small, one may use frequency-based approximation of the variance. The returns related to highly irregular trade volumes should be described by the market-based variance.

The empirical studies of statistical dependence between returns and trade volumes or returns and past values should consider market-based correlations of returns and squares of trade volumes or returns and squares of past values.

The lack of econometric observations and assessments of covariances of values and volumes of all transactions made in the economy reduces the number of economic-based predictable statistical moments of prices, returns, and macroeconomic variables by the first two. That limits the accuracy of economic-founded forecasts by the accuracy of Gaussian approximations.

To improve reliability of their forecasts and adjust large macroeconomic models like BlackRock's Aladdin, JP Morgan, and the U.S. Fed with reality determined by the impact of random trade volumes the developers may use the market-based variances.